\documentstyle[12pt,axodraw]{article}
\newcommand{\nc}{\newcommand}
\nc{\renc}{\renewcommand}

\nc{\half}{{\textstyle{1\over2}}}

\nc{\etal}{\mbox{\it et al. }}
\nc{\ie}{{\it i.e.}}
\nc{\eg}{{\it e.g.}}

\renc{\thefootnote}{\arabic{footnote}}
\nc{\capt}[1]{{\bf Figure.} {\small\sl #1}}

\nc{\eqs}[2]{\mbox{Eqs.~(\ref{#1},\,\ref{#2})}}
\nc{\eq}[1]{\mbox{Eq.~(\ref{#1})}}

\nc{\figs}[2]{\mbox{Figs.~(\ref{#1},\,\ref{#2})}}
\nc{\fig}[1]{\mbox{Fig.~(\ref{#1})}}

\nc{\tag}[1]{\label{#1} \marginpar{{\footnotesize #1}}}
\nc{\mtag}[1]{\label{#1} \mbox{\marginpar{{\footnotesize #1}}}}

\def\bort#1{ }
\renc{\baselinestretch}{1.2}
\newlength{\overeqskip}
\newlength{\undereqskip}
\nc{\be}[1]{\begin{equation} \mbox{$\label{#1}$}}
\nc{\bea}[1]{\begin{eqnarray} \mbox{$\label{#1}$}}
\nc{\Section}[2]{\section{#2}\label{#1}}
\nc{\Subsection}[2]{\subsection{\sc #2}\label{#1}}
\nc{\Bibitem}[1]{\bibitem{#1}}
\nc{\Label}[1]{\label{#1}}

\nc{\eea}{\vspace{\undereqskip}\end{eqnarray}}
\nc{\ee}{\vspace{\undereqskip}\end{equation}}
\nc{\bdm}{\begin{displaymath}}
\nc{\edm}{\end{displaymath}}
\nc{\dpsty}{\displaystyle}
\nc{\bc}{\begin{center}}
\nc{\ec}{\end{center}}
\nc{\ba}{\begin{array}}
\nc{\ea}{\end{array}}
\nc{\bab}{\begin{abstract}}
\nc{\eab}{\end{abstract}}
\nc{\btab}{\begin{tabular}}
\nc{\etab}{\end{tabular}}
\nc{\bit}{\begin{itemize}}
\nc{\eit}{\end{itemize}}
\nc{\ben}{\begin{enumerate}}
\nc{\een}{\end{enumerate}}
\nc{\bfig}{\begin{figure}}
\nc{\efig}{\end{figure}}
\nc{\figcap}[1]{\begin{quote}\refstepcounter{figure}
        {\bf Figure \thefigure}: {\small #1}\end{quote}}
\def\simleq{\; \raise0.3ex\hbox{$<$\kern-0.75em
      \raise-1.1ex\hbox{$\sim$}}\; }
\def\simgeq{\; \raise0.3ex\hbox{$>$\kern-0.75em
      \raise-1.1ex\hbox{$\sim$}}\; }
\nc{\arreq}{&\!=\!&}
\nc{\arrmi}{&\!-\!&}
\nc{\arrpl}{&\!+\!&}
\nc{\arrap}{&\!\!\!\approx\!\!\!&}
\nc{\non}{\nonumber\\*}
\nc{\align}{\!\!\!\!\!\!\!\!&&}

\def\lsim{\; \raise0.3ex\hbox{$<$\kern-0.75em
      \raise-1.1ex\hbox{$\sim$}}\; }
\def\gsim{\; \raise0.3ex\hbox{$>$\kern-0.75em
      \raise-1.1ex\hbox{$\sim$}}\; }
\nc{\DOT}{\hspace{-0.08in}{\bf .}\hspace{0.1in}}
\nc{\Laada}{\hbox {$\sqcap$ \kern -1em $\sqcup$}}
\nc\loota{{\scriptstyle\sqcap\kern-0.55em\hbox{$\scriptstyle\sqcup$}}}
\nc\Loota{{\sqcap\kern-0.65em\hbox{$\sqcup$}}}
\nc\laada{\Loota}
\nc{\qed}{\hskip 3em \hbox{\BOX} \vskip 2ex}

\def\Im{{\rm Im}\hskip2pt}
\nc{\real}{{\rm I \! R}}
\nc{\Z}{{\sf Z \!\!\! Z}}
\nc{\complex}{{\rm C\!\!\! {\sf I}\,\,}}
\def\bigid{\leavevmode\hbox{\small1\kern-3.8pt\normalsize1}}
\def\id{\leavevmode\hbox{\small1\kern-3.3pt\normalsize1}}
\nc{\slask}{\!\!\!/}
\nc{\bis}{{\prime\prime}}
\nc{\pa}{\partial}
\nc{\na}{\nabla}
\nc{\ra}{\rangle}
\nc{\la}{\langle}
\def\>{\rangle}
\def\<{\langle}
\nc{\goto}{\rightarrow}
\nc{\swap}{\leftrightarrow}

\nc{\EE}[1]{ \mbox{$\cdot10^{#1}$} }
\nc{\abs}[1]{\left|#1\right|}
\nc{\at}[2]{\left.#1\right|_{#2}}
\nc{\norm}[1]{\|#1\|}
\nc{\abscut}[2]{\Abs{#1}_{\scriptscriptstyle#2}}
\nc{\vek}[1]{{\rm\bf #1}}
\nc{\integral}[2]{\int\limits_{#1}^{#2}}
\nc{\inv}[1]{\frac{1}{#1}}
\def\d#1#2{d\,^{#1}#2\,}
\def\dbar#1#2{\frac{d\,^{#1}#2}{(2\pi)^{#1}}}

\nc{\dx}[1]{d\,^{#1}x}
\nc{\dy}[1]{d\,^{#1}y}
\nc{\dz}[1]{d\,^{#1}z}
\nc{\dl}[1]{\frac{d\,^{#1}l}{(2\pi)^{#1}}}
\nc{\dk}[1]{\frac{d\,^{#1}k}{(2\pi)^{#1}}}
\nc{\dq}[1]{\frac{d\,^{#1}q}{(2\pi)^{#1}}}

\nc{\cc}{\mbox{$c.c.$ }}
\nc{\hc}{\mbox{$h.c.$ }}
\nc{\cf}{cf.\ }
\nc{\erfc}{{\rm erfc}}
\nc{\Tr}{{\rm Tr\,}}
\nc{\tr}{{\rm tr\,}}
\nc{\pol}{{\rm pol}}
\nc{\sign}{{\rm sign}}
\nc{\bfT}{{\bf T }}

\nc{\cA}{{\cal A}}
\nc{\cB}{{\cal B}}
\nc{\cD}{{\cal D}}
\nc{\cE}{{\cal E}}
\nc{\cG}{{\cal G}}
\nc{\cH}{{\cal H}}
\nc{\cL}{{\cal L}}
\def\cM{{\cal M}}
\nc{\cO}{{\cal O}}
\nc{\cT}{{\cal T}}
\nc{\cN}{{\cal N}}
\nc{\rvac}[1]{|{\cal O}#1\rangle}
\nc{\lvac}[1]{\langle{\cal O}#1|}
\nc{\rvacb}[1]{|{\cal O}_\beta #1\rangle}
\nc{\lvacb}[1]{\langle{\cal O}_\beta #1 |}
\nc{\bb}{\bar{\beta}}
\nc{\bt}{\tilde{\beta}}
\nc{\ctH}{\tilde{\cal H}}
\nc{\chH}{\hat{\cal H}}
\nc{\1}{\aa}
\nc{\2}{\"{a}}
\nc{\3}{\"{o}}
\nc{\4}{\AA}
\nc{\5}{\"{A}}
\nc{\6}{\"{O}}
\nc{\al}{\alpha}
\nc{\g}{\gamma}
\nc{\Del}{\Delta}
\nc{\e}{\epsilon}
\nc{\eps}{\epsilon}
\nc{\lam}{\lambda}
\nc{\om}{\omega}
\nc{\Om}{\Omega}
\nc{\ve}{\varepsilon}
\nc{\mn}{{\mu\nu}}
\nc{\vp}{\varphi}

\nc{\advp}[3]{{\it  Adv.\ in\ Phys.\ }{{\bf #1} {(#2)} {#3}}}
\nc{\annp}[3]{{\it  Ann.\ Phys.\ (N.Y.)\ }{{\bf #1} {(#2)} {#3}}}
\nc{\apl}[3]{{\it  Appl. Phys. Lett. }{{\bf #1} {(#2)} {#3}}}
\nc{\apj}[3]{{\it  Ap.\ J.\ }{{\bf #1} {(#2)} {#3}}}
\nc{\apjl}[3]{{\it  Ap.\ J.\ Lett.\ }{{\bf #1} {(#2)} {#3}}}
\nc{\app}[3]{{\it Astropart.\ Phys.\ }{{\bf #1} {(#2)} {#3}}}
\def\cjp#1#2#3{{\it  Can.\ J.\ Phys.\ }{{\bf #1} {(#2)} {#3}}}
\nc{\cmp}[3]{{\it  Comm.\ Math.\ Phys.\ }{{ \bf #1} {(#2)} {#3}}}
\nc{\cqg}[3]{{\it  Class.\ Quant.\ Grav.\ }{{\bf #1} {(#2)} {#3}}}
\nc{\epl}[3]{{\it  Europhys.\ Lett.\ }{{\bf #1} {(#2)} {#3}}}
\nc{\ijmp}[3]{{\it Int.\ J.\ Mod.\ Phys.\ }{{\bf #1} {(#2)} {#3}}}
\nc{\ijtp}[3]{{\it Int.\ J.\ Theor.\ Phys.\ }{{\bf #1} {(#2)} {#3}}}
\nc{\jmp}[3]{{\it  J.\ Math.\ Phys.\ }{{ \bf #1} {(#2)} {#3}}}
\nc{\jpa}[3]{{\it  J.\ Phys.\ A\ }{{\bf #1} {(#2)} {#3}}}
\nc{\jpc}[3]{{\it  J.\ Phys.\ C\ }{{\bf #1} {(#2)} {#3}}}
\nc{\jap}[3]{{\it J.\ Appl.\ Phys.\ }{{\bf #1} {(#2)} {#3}}}
\nc{\jpsj}[3]{{\it J.\ Phys.\ Soc.\ Japan\ }{{\bf #1} {(#2)} {#3}}}
\nc{\lmp}[3]{{\it Lett.\ Math.\ Phys.\ }{{\bf #1} {(#2)} {#3}}}
\nc{\mpl}[3]{{\it  Mod.\ Phys.\ Lett.\ }{{\bf #1} {(#2)} {#3}}}
\nc{\ncim}[3]{{\it  Nuov.\ Cim.\ }{{\bf #1} {(#2)} {#3}}}
\nc{\np}[3]{{\it  Nucl.\ Phys.\ }{{\bf #1} {(#2)} {#3}}}
\nc{\pr}[3]{{\it Phys.\ Rev.\ }{{\bf #1} {(#2)} {#3}}}
\nc{\pra}[3]{{\it  Phys.\ Rev.\ A\ }{{\bf #1} {(#2)} {#3}}}
\nc{\prb}[3]{{\it  Phys.\ Rev.\ B\ }{{{\bf #1} {(#2)} {#3}}}}
\nc{\prc}[3]{{\it  Phys.\ Rev.\ C\ }{{\bf #1} {(#2)} {#3}}}
\nc{\prd}[3]{{\it  Phys.\ Rev.\ D\ }{{\bf #1} {(#2)} {#3}}}
\nc{\prl}[3]{{\it Phys.\ Rev.\ Lett.\ }{{\bf #1} {(#2)} {#3}}}
\nc{\pl}[3]{{\it  Phys.\ Lett.\ }{{\bf #1} {(#2)} {#3}}}
\nc{\prep}[3]{{\it Phys\. Rep.\ }{{\bf #1} {(#2)} {#3}}}
\nc{\prsl}[3]{{\it Proc.\ R.\ Soc.\ London\ }{{\bf #1} {(#2)} {#3}}}
\nc{\ptp}[3]{{\it  Prog.\ Theor.\ Phys.\ }{{\bf #1} {(#2)} {#3}}}
\nc{\ptps}[3]{{\it  Prog\ Theor.\ Phys.\ suppl.\ }{{\bf #1} {(#2)} {#3}}}
\nc{\physa}[3]{{\it  Physica\ A\ }{{\bf #1} {(#2)} {#3}}}
\nc{\physb}[3]{{\it  Physica\ B\ }{{\bf #1} {(#2)} {#3}}}
\nc{\phys}[3]{{\it Physica\ }{{\bf #1} {(#2)} {#3}}}
\nc{\rmp}[3]{{\it  Rev.\ Mod.\ Phys.\ }{{\bf #1} {(#2)} {#3}}}
\nc{\rpp}[3]{{\it Rep.\ Prog.\ Phys.\ }{{\bf #1} {(#2)} {#3}}}
\nc{\sjnp}[3]{{\it Sov.\ J.\ Nucl.\ Phys.\ }{{\bf #1} {(#2)} {#3}}}
\nc{\spjetp}[3]{{\it Sov.\ Phys.\ JETP\ }{{\bf #1} {(#2)} {#3}}}
\nc{\yf}[3]{{\it Yad.\ Fiz.\ }{{\bf #1} {(#2)} {#3}}}
\nc{\zetp}[3]{{\it Zh.\ Eksp.\ Teor.\ Fiz.\  }{{\bf #1}  {(#2)} {#3}}}
\nc{\zp}[3]{{\it Z.\ Phys.\ }{{\bf #1} {(#2)} {#3}}}
\nc{\ibid}[3]{{\sl ibid.\ }{{\bf #1} {#2} {#3}}}
\nc{\rf}[1]{(\ref{#1})}
\nc{\nn}{\nonumber \\*}
\nc{\bfB}{\bf{B}}
\nc{\bfv}{\bf{v}}
\nc{\bfx}{\bf{x}}
\nc{\bfy}{\bf{y}}
\nc{\vx}{\vec{x}}
\nc{\vy}{\vec{y}}
\nc{\oB}{\overline{B}}
\nc{\oI}{\overline{I}}
\nc{\oR}{\overline{R}}
\nc{\rar}{\rightarrow}
\nc{\ti}{\times}
\nc{\slsh}{\hskip-5pt/}
\nc{\sm}{Standard~Model~}
\nc{\MP}{M_{\rm Pl}}
\nc{\tp}{t_{\rm Pl}}
\nc{\ave}{\bar{E}}

\nc{\eff}{{\rm eff}}
\nc{\kk}{\vek{k}}
\nc{\pp}{{\rm p}}
\nc{\ga}{g_{a\gamma}}
\nc{\vv}{\\}
\nc{\eee}{{\bf E}}
\nc{\bbb}{{\bf B}}
\nc{\qcd}{T_{\rm QCD}}
\def\vec#1{{\bf #1}}
\def\tt{{\tilde{t}}}
\def\Ht{{\tilde{H}}}
\def\Wt{{\tilde{W}}}
\def\Htpm{{\tilde{H}^\pm}}
\begin{document}

{\title{\vskip-2truecm{\hfill {{\small HIP-1998-64/TH\\
\hfill SUITP-98-18 \\
\hfill TURKU-FL/P30-98 \\
\hfill CERN-TH/98-313\\}}}
{\bf Damping rates in the MSSM and electroweak baryogenesis}}

{\author{
{\sc Per Elmfors$^{1}$}\\
{\sl\small Department of Physics, Stockholm University,
Box 6730, S-113 85 Stockholm,
Sweden} \\
{\sc Kari Enqvist$^{2}$}\\
{\sl\small Department of Physics and Helsinki Institute of Physics} \\
{\sl\small P.O. Box 9,
FIN-00014 University of Helsinki,
Finland} \\
{\sc Antonio Riotto$^{3}$}\\
{\sl\small Theory Division, CERN, CH-1211 Geneva 23, Switzerland. }
\\
{\sc Iiro Vilja$^{4}$}\\
{\sl\small Department of Physics, University of Turku, FIN-20014 Turku,
Finland}
}}
\maketitle
\vspace{0.5cm}

\begin{abstract}
\noindent
We present an analysis of the
thermalization rate of Higgsinos and winos based on the 
imaginary part of the two-point Green function in
the {\it unbroken} phase of the MSSM. We use improved propagators  
including resummation of hard thermal loops and  the thermalization  
rate is computed at the one-loop level in the high temperature  
approximation. We find that the damping is typically dominated by  
scattering with gauge bosons, resulting in a damping rate of about 
$\gamma_{\Ht}\simeq 0.025T$, $\gamma_{\Wt}\simeq 0.065T$.
The contribution from scattering with scalars is relatively
small. Implications for baryogenesis are also discussed.

\end{abstract}
\vfil
\footnoterule
{\small $^1$elmfors@physto.se.};
{\small $^2$enqvist@pcu.helsinki.fi.};
{\small $^3$riotto@nxth04.cern.ch. On leave of absence from Theoretical Physics
Department, University of Oxford, UK.};
{\small $^4$vilja@newton.tfy.utu.fi.}
\thispagestyle{empty}
\newpage
\setcounter{page}{1}


The presence of  unsuppressed
baryon number violating processes at high temperatures within
the Standard Model (SM) of weak interactions makes the
generation of the baryon number at the electroweak scale an appealing scenario
\cite{reviews}. The baryon number violating processes    also impose severe
constraints on
models where the baryon asymmetry is created at energy scales much higher than
the electroweak scale \cite{anomaly}. Unfortunately, the electroweak phase
transition  is too weak  in the SM \cite{transition}
(although the existence of a primordial hypermagnetic field
could improve the situation \cite{eek}). This   means that the
baryon asymmetry
generated during the transition would subsequently be  erased by unsuppressed
sphaleron transitions in the broken phase.
 The most promising and
well-motivated framework for electroweak baryogenesis beyond the SM  seems to
be supersymmetry (SUSY).  Electroweak
baryogenesis in the framework of the Minimal Supersymmetric Standard Model
(MSSM) has  attracted much attention in the past years, with
particular emphasis on the strength of the phase transition
{}~\cite{early1} and
the mechanism of baryon number generation 
\cite{cpr,riottoprd,nelson,noi,higgs,ck}.

Recent  analytical \cite{r1,r2} and  lattice
computations  \cite{r3} have  revealed   that the phase transition can be
sufficiently strongly
first order if  the
ratio of the vacuum expectation values of the two neutral Higgses $\tan\beta$
is smaller than $\sim 4$. Moreover, taking into account all the experimental
bounds
including those coming from the requirement of avoiding dangerous
 color breaking minima,  the lightest Higgs boson should be  lighter than about
 $105$ GeV,
 while the right-handed stop mass might  be close to the present experimental
bound and should
 be smaller than, or of the order of, the top quark mass \cite{r2}.

Moreover, the MSSM contains additional sources
of CP-violation  besides the CKM matrix phase.
However, an acceptable baryon asymmetry from the stop sector
may only be generated through a delicate balance between the values
of the different soft supersymmetry breaking parameters contributing
to the stop mixing parameter, and their associated CP-violating
phases \cite{noi}. As a result, the contribution to the final baryon asymmetry
from the stop sector turns out to be negligible.   On the other hand, charginos
and neutralinos may be responsible for the observed baryon asymmetry
in the MSSM\cite{noi,ck,mv}.
If the strength of the
 electroweak phase transition is enhanced by the presence of some
new degrees of freedom beyond the ones contained in the MSSM, {\it
e.g.} some extra standard model gauge singlets,
 light stops (predominantly the
right-handed ones) and charginos/neutralinos are expected to
give quantitatively the
same contribution to the final baryon asymmetry.

CP-violating sources in supersymmetric baryogenesis  are
more easily built up if the degrees of freedom  in  the
stop and gaugino/neutralino
sectors are nearly degenerate in mass \cite{noi}.
\bort{
This is due to a large
enhancement of the computed baryon asymmetry for these values
of the parameters.
}
This resonant behaviour is
associated with the
possibility of absorption (or emission) of Higgs quanta by the
propagating supersymmetric particles. For momenta of the order of the critical
temperature, this can only take place when, for instance,  the Higgsinos and
gauginos do not differ too much in mass.
By using the Uncertainty Principle, it is easy to understand that the
width of this resonance is expected
to be proportional to the thermalization rate  of the particles giving rise to
the baryon asymmetry \cite{noi}.

 From all these considerations, it is  clear that the computation of the
thermalization
rate of
the particles responsible for supersymmetric baryogenesis represents a
necessary and crucial step towards  the final computation of the baryon number.
A detailed
computation of the   thermalization rate of the  right-handed stop from the
imaginary
 part of the two-point Green function has been recently performed in
\cite{therm} by making use of improved propagators and including
 resummation of hard thermal loops. The thermalization rate has been  computed
exactly at the
 one-loop level in the high temperature approximation as a function of the
plasma right-handed stop mass
 $m_{\widetilde{t}_R}(T)$ and an  estimate for the magnitude of the
 two-loop contributions which dominate the rate for small
$m_{\widetilde{t}_R}(T)$ was also given.
If  $m_{\widetilde{t}_R}(T)\gsim  T$, the thermalization  is dictated by the
one-loop thermal decay rate which can be larger than $T$ \cite{therm}. For
smaller values of   $m_{\widetilde{t}_R}(T)$, when the
 thermalization is dominated by two-loop effects (i.e. scattering),
$\Gamma_{\widetilde{t}_R}$ may be as large as $10^{-3} T$ \cite{therm}.
 With such value, our derivative expansion is
perfectly justified since the wall thickness can span the range
$(10-100)/T$.

The goal of this paper is to present a computation of the
thermalization rate of  Higgsinos and winos
from the imaginary part of the two-point Green function in
the {\it unbroken} phase of the MSSM. We use improved propagators including
resummation of hard thermal loops and  the thermalization rate is computed at
the one-loop level in the high temperature approximation.

There are two types of diagram contributing to the damping rate of charginos
and neutralinos to one
loop. The internal loop has either a scalar or a vector boson, apart from
the fermion. We expect the leading processes to involve massless gauge
bosons, since these diagrams are IR divergent without plasma resummation.

Let us begin with  
performing an analysis for the charged
$\Ht^\pm$. We first calculate the gauge boson contribution (as
accurately as possible with the present understanding of the IR sensitivity)
and then compare it with a typical diagram involving a scalar (the stop). We
find that scattering with scalars provides typically much less 
damping.
The contribution from unbroken gauge bosons is in some sense the most
difficult diagram since it requires subtle resummation schemes. On the other
hand, this is also the most studied case and the result can be found in the
literature. The relevant diagram is
\newline
\begin{picture}(160,110)(-110,20)
        \SetScale{3.4}\setlength{\unitlength}{1mm}
        \Line(15,15)(50,15)
        \PhotonArc(32,15)(10,0,180){1}{8}
        \GCirc(22,15){3}{0.5}
        \GCirc(32,25){3}{0.5}
        \GCirc(42,15){3}{0.5}
        \GCirc(32,15){3}{0.5}
	\Text(10,18)[]{$\tilde{H}^\pm$}
	\Text(39,10)[]{$\tilde{H}^{\pm,0}$}
	\Text(39,37)[b]{$B$ or $W^{\pm,0}$}
\end{picture}
\newline
Depending on the assumptions of the mass and momentum of
the external fermion there are a number of different cases:
\bit
\item
{\bf Heavy $m\gg T$ fermion at rest $p=0$.}\\
In this case thermal fermions are not present but only the gauge bosons. The
result for the damping rate $\gamma$ is \cite{Pisarski89}
\be{mgtTrest}
	\gamma=\frac{g^2TC_f}{8\pi}~~,
\ee
Here $C_f$ denotes the
quadratic Casimir invariant of the fermion representation.

\item
{\bf Massless fermion at rest $p=0$.}\\
This calculation is more involved than the heavy fermion case since one needs
to resum all propagators and vertices.
In \cite{BraatenP90}
this is done for a few  different gauge groups
with the result that
\be{mlrest}
	\gamma\equiv-\inv8\Im\tr[\g_0\Sigma(p_0=\cM ,p=0)]
	=a(N,N_f)\frac{g^2TC_f}{4\pi}~~,
\ee
where $a(3,2)=1.40$, $a(2,2)=1.45$ and $a(2,4)=1.57$.
The same quantity was calculated in \cite{BraatenP92}
 with the result $a(3,3)=1.43$.
A major problem with this calculation is that
the gauge independence of $\gamma$ has not been established.
With certain particular IR regularization schemes the result is gauge invariant
but depends on how the IR regulator is removed and 
the external particle put on-shell \cite{BaierKS92,Smilga97}.

\item
{\bf Moving $p>0$ massless fermion.}\\
There have been several papers concerning this case and most of them
conclude that $\gamma$ depends on the magnetic screening
\cite{LebedevS90,Smilga97,Pisarski93}.
For  abelian theories this is not an alternative so another resummation
scheme was presented
in \cite{IancuB96}  (apart from the usual HTL resummation) which resulted in a
non-exponential but finite decay of correlation functions. It is found that
excitation with $p\gg e^2 T$ (which is what we have) decay like
\be{nonexpdec}
	S(t)=e^{-i\omega_p t}\exp[-v\frac{e^2T}{4\pi} t \ln(\cM v t)]~~,
\ee
where again $v$ is the velocity and $\cM$ is the plasma mass. The
expectation value of the damping rate is to leading order
\be{mlmov}
	\gamma=v\frac{e^2T}{4\pi}\ln\left(\frac{4\pi\cM}{e^2T}\right)~~,
\ee
which has the same leading $e^2T\ln\inv e$ term as the heavy fermion
case. The same leading damping rate
has been obtained for the non-abelian plasma in
 \cite{FlechsigRS95,Pisarski93} by assuming the
existence of a magnetic mass of order $g^2T$ but the solution in
\cite{IancuB96} is independent of such assumptions. Notice also that
in \cite{IancuB96} the constant in $\gamma\sim e^2T(\ln\inv e+{\rm
const})$ was calculated.
\eit
 From the results above we conclude that, to the extent that it is
 calculable,  the damping rate of a light fermion due to
scattering with gauge bosons is approximately
$\gamma\simeq\frac{g^2T}{4\pi}C_f$ for both $p=0$ and $p>0$. The correction
factor $\ln(\frac{4\pi\cM}{g^2T})$ for moving fermions and the variation in
the factor $a(N,N_F)$ is at most of order 2 and does not affect the
conclusion significantly.

For Higgsinos, the gauge boson contributions reflect weak and
hypercharge symmetries. For the contribution from the weak gauge bosons
we obtain $\gamma \propto 3 g_2^2 T/(16 \pi) \simeq 0.0244 T$ ($g_2 = 0.640$).
The hypercharge contribution reads $ g_1^2 T/(16 \pi) \simeq 0.00122 T$
($g_1 = 0.247$) which is negligible compared to the weak contribution.
Thus we conclude that the gauge boson contribution to the damping rate of
$\Ht^\pm$ is
\be{gbc}
	\gamma_{\rm gb} \simeq 0.025 T
\ee
up to a factor 2 depending on non-perturbative effects through the magnetic
mass and variations in the factor $a(N,N_f)$. 

Let us now turn to scattering with scalars.
There are a number of diagrams where the external fermion scatter with
various scalars such as Higgs or squarks. They differ in structure depending
on whether the fermion is Dirac or Majorana and whether the external
Higgsino has a diagonal or off-diagonal mass term. However, the kinematics
and coupling constants are all similar. Moreover, they do not suffer from
the IR sensitivity of gauge boson scattering. Therefore, we shall only
calculate explicitly the simplest diagram which
is the stop-bottom loop  with an external $\Htpm_2$ where the
internal chiral
fermion has no Majorana mass and the external particle has only a
diagonal Dirac mass. Taking the notation from \cite{KobesS85} the
imaginary part is given by
\bea{ImSig}
	\Im\Sigma(P)&=&-\frac{h_\tt^2\sign(p_0)}{\sin 2\phi_P}
	\int\dbar4K\inv{2}\sin 2\phi_K \inv{2}\sinh 2\phi_{K-P}
	\nn &&
	2\pi \delta\biggl((K-P)^2-\cM_\tt^2\biggr)
	\inv2 (1-\g_5) 2\pi \cA_b(k_0,k)~~,
\eea
where
\bea{notation}
	\cA_b(K)&=&\inv{2\pi i}
	\left(S_b(k_0-i\eps,k)-S_b(k_0+i\eps,k)\right)~~,
	\nn
	S_b(K)&=&\frac{s(K)k_0\g_0-r(K)\vek{k}\cdot\vek{\g}}
	{s(K)^2 k_0^2-r(K)^2 k^2}~~,
	\nn
	s(K)&=&1-\frac{\cM_b^2}{2k_0 k}\ln\left(\frac{k_0+k}{k_0-k}\right)~~,
	\nn
	r(K)&=&1+\frac{\cM_b^2}{k^2}\left[1-\frac{k_0}{2k}
	\ln\left(\frac{k_0+k}{k_0-k}\right)\right]~~.
\eea
The notation is that $\cM_b$, $\cM_\tt$, $\cM_\Htpm$ and $\mu$ stands for the
thermal bottom, stop, Higgsino mass and the vacuum Higgsino mass,
respectively.
Since $s(K)$ and $r(K)$
have imaginary parts for $k_0^2<k^2$ there is a continuous
part of the spectral function $\cA_b$
apart from the $\delta$-functions. We shall
use the notation $s(k_0-i\epsilon,k)=s_R(K)+i\,s_I(K)$ and
similarly for $r(K)$.

We want to take the expectation value of $\Im\Sigma$ in the external state
of Higgsinos with thermally corrected masses. Since they have several
branches and a continuous spectral weight below the light cone
it is not possible to use a simple particle picture for the external states.
One straightforward way to include collective effects is to
multiply $\Im\Sigma$ with
the spectral function for $\Htpm$, i.e.
$\cA_\Htpm(P)$, and take the trace and integrate
over $p_0$. This gives the damping
for a Higgsino averaged over all states with a given momentum.
When the external states have several branches we have to be careful
not to overcount the degrees of freedom, but the spectral function
conveniently satisfies an exact sum rule which counts the number of
degrees of freedom.
Assuming that the external Higgsino mass is dominated by thermal effects we
can use \eq{mlrest} for the damping rate:
\be{inttrASig}
	\gamma_{b/\tt}=-\inv2\<\Im\Sigma\>(\vec{p})=-\inv 4\int_0^\infty
	dp_0 \tr[\cA_\Htpm(P) \Im\Sigma(P)]~~,
\ee
where we have divided  by a factor 2 to average over the spins for
incoming positive energy solutions.
Finally we should do the $K$ integral. For simplicity we choose $\vec{p}=0$
so that there is no continuous contribution from the external states.
Then the  angular part is easy so we
rewrite it as
\be{Kint}
	\int\d4K\delta((K-P)^2-\cM_\tt^2)
	=
	\inv{8\pi^3}\int_{-\infty}^\infty
	dk_0 \,k(k_0)\,\theta((k_0-p_0)^2-\cM_\tt^2)~~.
\ee
Finally, we end up with
\bea{final}
	&&\inv{4}\int_0^\infty dp_0 \tr[\cA_\Htpm \Im\Sigma]
	=
	-\frac{h_\tt^2}{32 \pi^2}\sum_{p_0^{(j)}>0}
	\frac{p_0^2}{p_0^2+\cM_\Htpm^2}
	\int dk_0\,k\,\theta((k_0-p_0)^2-\cM_\tt^2)
	\nn &&
	\times\sign(k_0-p_0)
	\left[\inv{e^{\beta k_0}+1}+\inv{e^{\beta (k_0-p_0)}-1}\right]
	\nn &&\times
	\left\{  \theta(k_0^2-k^2)\frac{(k_0 - \eta k)(k_0^2-k^2)}
		{4\cM_b^2(k_0^2- k^2-\cM_b^2)}
	s(K)k_02\pi\sum_i\delta(k_0-k_0^{(i)}(k))
	\right. \\ && \left.
	 +\theta(k^2-k_0^2)
	\frac{s_Ik_0
		[(s_R^2-s_I^2)k_0^2-(r_R^2-r_I^2)k^2]
	-2s_Rk_0[s_Rs_Ik_0^2-r_Rr_Ik^2]}
	{[(s_R^2-s_I^2)k_0^2-(r_R^2-r_I^2)k^2]^2
	+4[s_Rs_Ik_0^2-r_Rr_Ik^2]^2}
	\right\}~~,\nonumber
\eea
where $k=\sqrt{(k_0-p_0)^2+\cM_\tt}$.
This should be summed over the two solutions
$p_0^{(\pm)}=(\sqrt{\mu^2+4\cM_\Htpm^2}\pm \mu^2)/2$.
In first term of the curly bracket the
$\delta$-function is used for the $k_0$ integral so in practice one has to
find all simultaneous solutions to
\be{kk0}
	(k_0-p_0)^2-k^2-\cM_\tt^2=0~,~~{\rm and}\quad
	D_b(K)=0~~,
\ee
and use the values of $k_0$ and $k$ in the integrand. For the continuous part
it is in principle necessary to
integrate over all $k_0$ and for each value of $k_0$
and solve $k$ from $k=\sqrt{(k_0-p_0)^2-\cM_\tt^2}$. At the same time
there are the factors
$\theta((k_0-p_0)^2-\cM_\tt^2)\,\theta(k^2-k_0^2)$
that restrict the integration domain.

\begin{figure}[ht]
\leavevmode
\centering
\vspace*{80mm}
\begin{picture}(100,60)(0,0)
\includegraphics{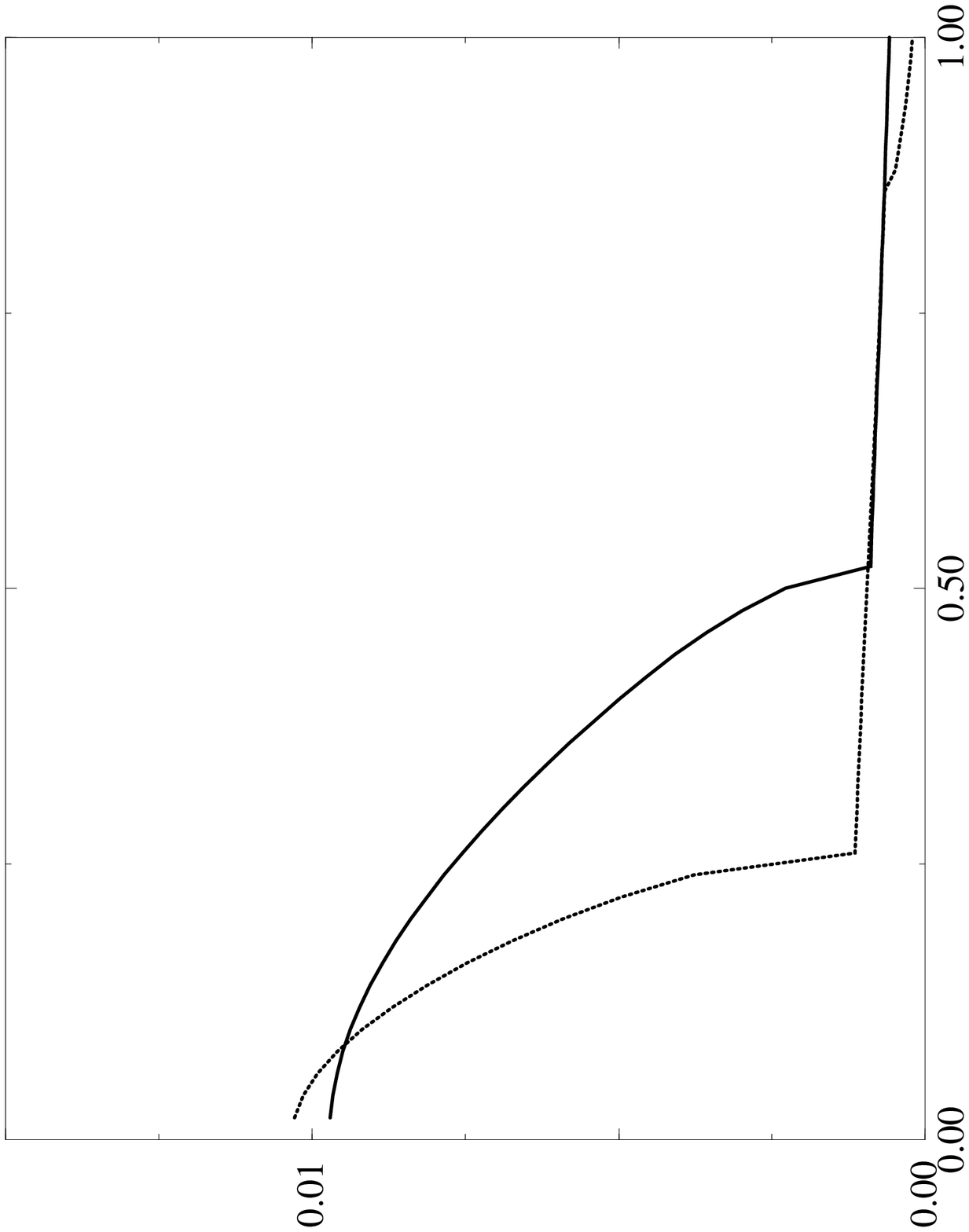}
   \put(-170,20){\large $\gamma_{b/\tt}$}
   \put(120,-215){\large $\cM_\tt$}
\end{picture}
\figcap{Scalar contribution to the Higgsino damping rate as a function of
right-handed stop mass in units of $T$. 
The different curves are for $p_0 = 1.0 T$ (solid line) and $p_0 = 0.75 T$
(dotted line).
\label{f:kuva}
}
\end{figure}

We have evaluated the damping rate for a Higgsino at rest
numerically using $h_\tt^2=1$ for the Higgsino Yukawa
coupling (see \fig{f:kuva}).
The scalar contribution to the thermalization rate (in the units of
temperature) is presented for
two different
values of the effective thermal Higgsino mass:
$p_0 = 1.0 T$ (solid line) and $p_0 = 0.75 T$
(dotted line), and for range of the right-handed stop mass form $0$ to $T$. It
shows that for large enough stop mass (${\cal M}_{\tilde t} \gsim 0.5 T$
for $p_0 = 1.0 T$ and ${\cal M}_{\tilde t} \gsim 0.25 T$
for $p_0 = 0.75 T$)
the main contribution to the damping rate comes from the (weak) gauge bosons.
For the lower end of the stop mass range, however, the scalar contribution
is considerably larger ($\gamma_{b/\tt}\simeq 0.01T$) but still smaller
than $\gamma_{\rm gb}$.

The aim of this paper has been to establish a reliable value for the damping
rate of Higgsinos in MSSM above the electroweak phase transition. Our claim
is that the damping is typically dominated by scattering with gauge bosons
and that the contribution from scattering with scalars is relatively
small. We found that the gauge boson contribution to the damping is given by
$\gamma_{\rm gb}\sim 0.025 T$. Scattering with stop/bottom gives a largest
contribution for small stop mass (see \fig{f:kuva}) $\gamma_{b/\tt}\simleq
0.01T $, but is much smaller for higher stop mass. Other scattering
processes (involving gauginos and scalars Higgses) can be expected to give
similar phase space integrals as the stop/bottom scattering but with gauge
coupling constants instead of $h_\tt$. We can therefore conclude that they
are relatively insignificant.
In a similar way we can estimate the damping rate of $\Ht^0$ and
$\tilde{W}$. Actually, neutral Higgsino damping rate coincides with the charged 
Higgsino rates, because we are working in the unbroken phase.
They also scatter with gauge bosons and we expect this to be
the dominant process. In that case the damping rates would be
$\gamma_{\Ht^0}\simeq \gamma_{gb} \simeq 0.025 T$ and 
$\gamma_{\tilde{W}}\simeq g_2^2T/2\pi=0.065T$.

We would like to reiterate why we only considered one single diagram including 
scattering with scalars. Even though all of them can
be calculated with high accuracy using resummed propagators there is no
reason why they should differ significantly numerically, apart from have
different overall coupling constants. On the other hand, scattering of
moving fermions 
with gauge bosons is highly non-trivial problem that can only be
estimated using assumption about the value of the non-perturbative magnetic
mass. Different estimates give qualitatively the same result but the
accuracy can of course not be guaranteed. (The situation is different in
abelian theories where there exists a resummation procedure controlling the
IR sensitivity \cite{IancuB96}.) It is, therefore, for the moment, 
no point in trying to
achieve higher accuracy in the scalar scattering diagrams which anyway are
subdominant.  The case is
different for $\tilde{B}^0$ which is gauge singlet and thus does not have
(one-loop) gauge boson contribution. However, the form of its scalar
contribution differs from the scalar contribution of the Higgsinos because
of the Majorana nature of $\tilde B^0$ \cite{maj}.

Let us briefly discuss the implications of our findings. As we already
mentioned, the precise  knowledge of the thermalization rate
of the supersymmetric particles is a key ingredient for the
computation of  the final baryon asymmetry. Sizeable decay rates of the  particles propagating in the plasma destroy  the quantum interference out of
which the   the CP-violating sources are built up  and therefore reduce the 
baryon asymmetry. Small decay rates, on the other side, are relevant when the 
particles reflecting off the advancing bubble wall  have
comparable masses and resonance effects show up  \cite{noi}.
In such a case, the thermalization  rates provide  the natural width of these
resonances and as the present calculation demonstrates, in supersymmetric
theories these depend in a complicated way on the particles involved
and their plasma masses. Moreover, as shown in \cite{riotto97,riottomore}, memory
effects
proportional to the damping rate of the Higgsinos play a crucial role in
determining the lowest  value of the CP-violating phases necessary to account
for the observed baryon asymmetry. In particular,  memory effects introduce an enhancement factor  in the final baryon asymmetry of the order of 
\begin{equation}
\label{new}
\left(\frac{\gamma_{\widetilde{H}}}{5\times 10^{-2}\:T}\right)\left(\frac{0.1}{v_\omega}\right)
\end{equation}
with respect to the results found in \cite{noi}. Here $v_\omega$ is the velocity of the bubble wall propagating in the plasma.  
Combining our findings 
with the analysis of Ref. \cite{riottomore} indicate that
the phase $\phi_\mu$ of the parameter $\mu$ satisfy the inequality
\begin{equation}
|\sin(\phi_\mu)|\gsim 5\times 10^{-2}\left(\frac{v_\omega}{0.1}\right).
\end{equation}
These small values of the phase $\phi_\mu$ are consistent with the constraints from the electric dipole moment of the neutron even if the squarks of the first and second generation have masses of the order of 300 GeV.

\vskip 1cm
\subsection*{Acknowledgments}
We would like to
thank NorFA for providing partial financial support from the NorFA
grant 96.15.053-O. P.~E. was financially supported by the Swedish Natural
Science Research Council under contract 10542-303, and K.~E. by the Academy
of Finland  under contract 101-35224.

\def\NPB#1#2#3{{\it Nucl. Phys.} {\bf B#1} (19#2) #3}
\def\PLB#1#2#3{{\it Phys. Lett.} {\bf B#1} (19#2) #3}
\def\PLBold#1#2#3{{\it Phys. Lett.} {\bf#1B} (19#2) #3}
\def\PRD#1#2#3{{\it Phys. Rev.} {\bf D#1} (19#2) #3 }
\def\PRL#1#2#3{{\it Phys. Rev. Lett.} {\bf#1} (19#2) #3}
\def\PRT#1#2#3{{\it Phys. Rep.} {\bf#1} (19#2) #3}
\def\ARAA#1#2#3{{\it Ann. Rev. Astron. Astrophys.} {\bf#1} (19#2) #3}
\def\ARNP#1#2#3{{\it Ann. Rev. Nucl. Part. Sci.} {\bf#1} (19#2) #3}
\def\MPL#1#2#3{{\it Mod. Phys. Lett.} {\bf #1} (19#2) #3}
\def\ZPC#1#2#3{{\it Zeit. f\"ur Physik} {\bf C#1} (19#2) #3}
\def\APJ#1#2#3{{\it Ap. J.} {\bf #1} (19#2) #3}
\def\AP#1#2#3{{\it Ann. Phys. } {\bf #1} (19#2) #3}
\def\RMP#1#2#3{{\it Rev. Mod. Phys. } {\bf #1} (19#2) #3}
\def\CMP#1#2#3{{\it Comm. Math. Phys. } {\bf #1} (19#2) #3}

\end{document}